# Voltage-Dependent Differential Conductance (d$I$/d$V$) Imaging of a Polymer:Fullerene Bulk-Heterojunction

*Goutam Paul, Biswajit Kundu, and Amlan J. Pal\**

Department of Solid State Physics, Indian Association for the Cultivation of Science, Jadavpur, Kolkata 700032, India

E-mail: sspajp@iacs.res.in.

Abstract

With scanning tunneling spectroscopy (STS), we probed differential conductance (d$I$/d$V$) images of P3HT:PCBM bulk-heterojunctions (BHJs). Since the materials are seen "energetically" in d$I$/d$V$ images, the imaging process provides opportunities to view the nano-domains of the components in the BHJ. The images were recorded at different voltages to bring out the interface region between the two materials as well. The density of states (DOS) spectra of pristine materials provided location of energy levels of the polymer and the fullerene in forming energy diagram with a type-II alignment from the view-point of charge carriers. DOS spectra recorded in the P3HT:PCBM heterojunction in addition yielded energies that deviated from those of the components indicating bending of energy levels at the interface region.

Electronic and opto-electronic devices are mostly based on junctions between at least a pair of semiconductors [1]. The semiconducting materials can be organic or inorganic or even a combination of two, that is, hybrid in nature. In all the cases, it is often extremely important to view the individual materials in a junction, since domain size and domain purity and thereby interface region are very crucial in achieving efficient device performance [1,2]. The viewing can become critical when physical dimension of the materials is restricted down to nanometer regime [3].

Semiconductors in their lower-dimensional form can in general be identified through their energy levels. The energies can be measured in a localized manner through scanning tunneling spectroscopy (STS) [4-7]. The spectroscopy provides differential tunnel conductance ($dI/dV$) that has correspondence to the semiconductor's density of states (DOS) at the point of measurement. Band-edges or energy levels with respect to Fermi energy can therefore be derived that bear the signature of the semiconductor even when the material is in the nanoscale regime [7-9].

It is needless to mention that P3HT:PCBM ought to be discussed when a bulk-heterojunction (BHJ) is referred. P3HT:PCBM is probably the most widely studied BHJ in the field of conjugated organics [1]. A heterojunction like such can be "energetically-imaged" through scanning of $dI/dV$ over an area. In a solitary report in this direction, $dI/dV$ images were recorded in a 100 nm thick P3HT:PCBM BHJ through cross-sectional STS measurements to look for interpenetrating percolating phases [7]. For completeness, P3HT and PCBM refer to poly(3-hexylthiophene-2,5-diyl) and [6,6]-phenyl-$C_{61}$-butyric acid methyl ester, respectively.

Since DOS of a semiconductor has a strong bias-dependence [10-13], $dI/dV$ images should in principle provide more than the phases of BHJs if they are recorded at different biases.

Voltage-dependent d*I*/d*V* images may bring out interface regions as well that may possess a shift in energy levels. In this work, we therefore have taken up the well-documented P3HT:PCBM BHJ to record voltage-dependent d*I*/d*V* images and analyzed them in depth.

*Materials.* While regioregular P3HT was purchased from Sigma Aldrich Chemical Company, PCBM was procured from SES Research. Both the components of the BHJ were used without further purification. Au(111) substrates used for STS measurements were purchased from M/s PHASIS Sàrl, Switzerland.

*Film Formation.* P3HT and PCBM were handled in a globe-box. 14 mg of P3HT was first dissolved in 1 mL dichlorobenzene through overnight stirring at 60 °C. To form ultrathin films of P3HT, the solution was spun on Au(111) substrates at 6000 rpm for 60 s. The films were then allowed to dry in inert environment for at least 30 min. To form PCBM films, a solution containing 14 mg of PCBM in 1 mL dichlorobenzene was used. For thin-films of P3HT:PCBM BHJ, a solution containing 20 mg of P3HT and 30 mg of PCBM in 3.5 mL dichlorobenzene was used, so that the components have about 1:1 v/v in the BHJ. The film-formation and drying process for PCBM and BHJ remained the same. For STS studies, the films were quickly transferred to the load-lock chamber of a scanning tunneling microscope (STM) and then to the main chamber following usual steps and protocol.

*STS Studies.* The ultrathin films of P3HT, PCBM, and P3HT:PCBM BHJs were characterized in a Pan-style ultrahigh vacuum STM (UHV-STM) of M/s RHK Technologies Inc. While the pressure of the chamber was $1.2 \times 10^{-10}$ Torr, temperature of the substrate and the tip both was maintained at 80 K.

For STS studies, the Pt:Ir tip was "approached" till a set-point of 0.1 nA at 2.0 V was achieved through a feedback loop. Differential tunnel conductance (d*I*/d*V*) or DOS spectra of the materials were recorded using a standard lock-in technique with a modulation frequency of 951 Hz and a rms voltage of 20 mV. The DC bias applied to the substrate was swept between -2.5 to 2.5 V. d$I$/d$V$ "images" of the films were formed through scanning of d$I$/d$V$ over a small region. Such images were recorded at range of DC voltages of both bias-directions. It may be stated that in a d$I$/d$V$ image, the material with a higher DOS at that voltage becomes visible brightly.

We first aimed to form an energy-level diagram of the P3HT:PCBM heterojunction. To do so, we recorded differential tunnel conductance spectra of the individual materials in their ultrathin forms. Typical spectra of P3HT and of PCBM are shown in Fig. 1. Since a d$I$/d$V$ spectrum has a correspondence to the DOS of the semiconductor and since the bias was applied to the substrate electrode, the peak (in a d$I$/d$V$ spectrum) in the positive voltage region closest to 0 V indicated the location of LUMO inferring injection of electrons to the conjugated organics. In the negative voltage, the peak similarly provided location of HOMO from where electrons could be withdrawn. Since STS is an extremely localized mode of measurement, it is imperative to record spectrum at many points on an ultrathin film and to locate energy levels from each spectrum. Histogram of HOMO and LUMO energies are then plotted to obtain the actual levels of a semiconductor (Fig. 2). The gap between the energy-levels in the two materials matched well with the values conventionally used in the literature [14,15]. The gap is expectedly higher than the respective material's optical gap, which takes exciton binding energy into account. The energy distributions, which are symmetric and Gaussian in nature, may have arisen out of defect or disordered states. The distribution of P3HT's energy levels appeared to be broader than that of PCBM implying possibility of more defect/disordered phases in the polymer as compared to the fullerene.

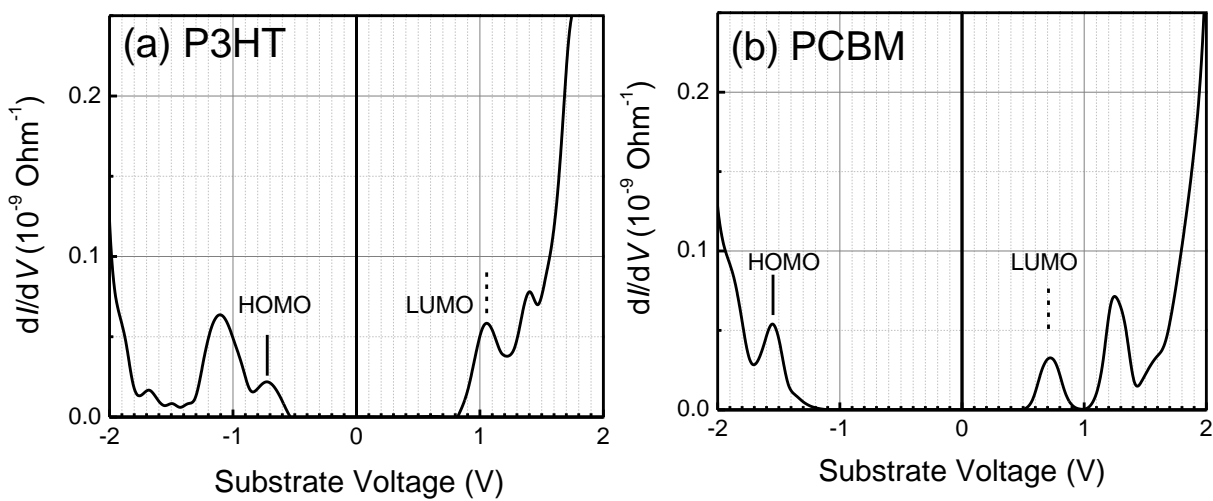

FIG. 1. Differential tunnel conductance versus voltage characteristics of an ultrathin-film of (a) P3HT and (b) PCBM.

We then proceeded to record DOS spectra of the P3HT:PCBM bulk-heterojunction. Here also, we recorded many spectra on an ultrathin film of the BHJ. Since STM topography does not provide any information about the material at the point of measurement, DOS spectra could be recorded without any prejudice. The material at the point of measurement can be identified only from an analysis of the DOS spectrum and peaks thereof. The figure prima facie brings HOMO and LUMO levels of the two components in the BHJ. The peaks in the histograms referring energy levels of P3HT and PCBM did not show any shift when compared to the energies of respective materials in their pristine films (Figs. 2a and 2b). The d$I$/d$V$ measurements and histograms of HOMO and LUMO levels of P3HT:PCBM BHJ (Fig. 2c) moreover allowed us to derive distributions in the energies of the components in a BHJ.

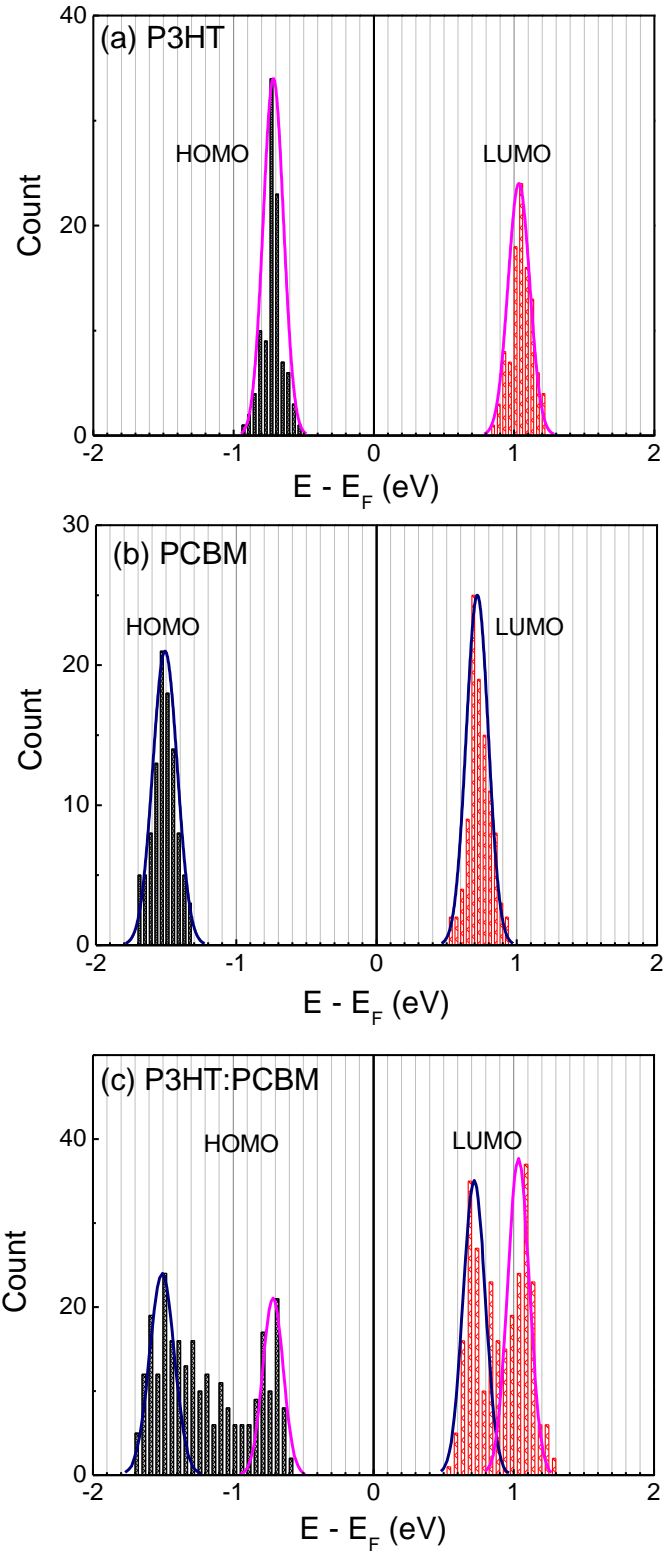

FIG. 2. Histogram of HOMO and LUMO energies as derived from many d$I$/d$V$ spectra of an ultrathin-film of (a) P3HT, (b) PCBM, and (c) P3HT:PCBM BHJ. The lines show distribution of energies. In (c), normalized distribution of the components' histograms have been added instead.

The histograms of energies in the BHJ could be found to be more than a sum of the histograms of the components' energies. Their comparison reveals that there were reasonable counts at energies other than the levels of the components. That is, when we appended normalized distribution of the components' energies in Fig. 2c (of BHJ), we find that a reasonable number of sticks appeared outside the distributions. Interestingly, the histograms of energies in the BHJ did not extend to both energy-directions. Each level of the two materials extended towards each other. To be specific, the distribution of P3HT's HOMO extended towards the higher (magnitude) of energy. The one of PCBM's HOMO, on the other hand, extended towards the lower-energy direction. Distributions of the LUMOs in the BHJ were similarly asymmetric with one wider wing; the LUMO also extended towards each other. The broadenings in the LUMO levels were although lesser than that of the HOMOs.

A broadening in energy distribution first of all can occur due to formation of domains, at which the material's energy levels are affected due to an interface with the other material. That is, PCBM at the periphery of P3HT's domains may affect the energy levels of P3HT and vice versa. In fact, the size of domains when controlled through additives has recently been shown to widen distribution of energy levels [16]. Here, the asymmetric energy distribution in a BHJ implied energy-level-bending at the interface. The nature of asymmetry, that is, each set of levels of the two materials extending towards each other, adds credence to such energy-level-bending. When we draw energy-level diagram of a P3HT:PCBM BHJ, such a bending of the levels at the interface can be seen clearly (Fig. 3).

In BHJs, it is often argued that the materials deep inside the domains form crystallites whereas amorphous phases are formed at the periphery due to presence of or intermixing with the other material [17-19]. Since the amorphous phases possess a higher energy than the crystalline one, the distribution of energy levels may hence become broadened. Such a widening in the width of energy-distribution would however be towards the high energy-direction only. That is, distribution of both the HOMO levels would be extended towards the negative energy, whereas the LUMOs should span towards the positive energy side. In our studies, the widening in the distribution of energy levels (Fig. 2c) were however asymmetric. This asymmetric broadening must hence have occurred due to energy-level-bending at the interface between P3HT and PCBM. The quantum of broadening should depend on the offset in the energy levels of each kind. Since HOMO-offset was higher than that of the LUMOs, the asymmetric broadening of energy-levels was predominant for the HOMO levels.

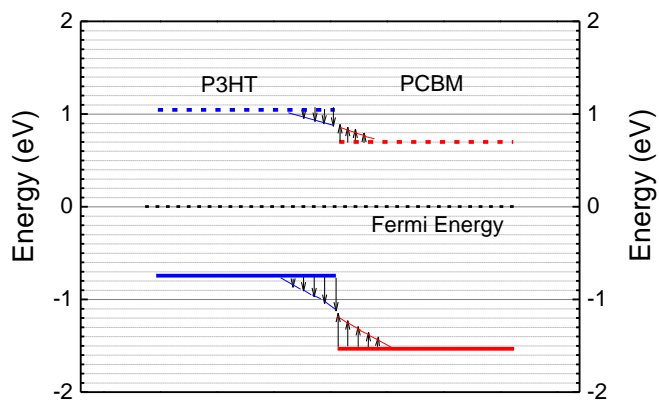

FIG. 3. Energy level diagram of P3HT:PCBM BHJ. The arrows indicate possible energy-level-bending near the interface between the two materials.

With histogram of the two components' HOMO and LUMO energies (Figs. 2a and 2b), we have drawn energy-diagram at the P3HT:PCBM junction (Fig. 3). The materials expectedly formed a type–II energy-alignment at the interface. In true sense, energy levels of the components in isolation were used in forming the diagram. At the interface, one requires to consider the energy-level bending that we just discussed. The DOS measurements hence allowed us to draw the energy-level diagram from the perspective of charge carriers in a device. On the basis of this diagram, voltage dependence of d$I$/d$V$ images can be analyzed to identify and view domains of polymer and fullerene in a bulk-heterojunction

It is important to know if the distribution of HOMO and LUMO levels were correlated. That is, if the shift in one level was accompanied by a similar shift of the other one. Such a correlated shift would support a bending of energy levels of one material due to other one's influence in a bulk-heterojunction. To make such studies, we have measured d$I$/d$V$ spectra along a line of a BHJ. While the STM topography (Fig. 4a) expectedly did not provide any information on the material, peak positions at the d$I$/d$V$ spectra allowed us to identify the material at each point of measurement.

When we analyzed the spectra obtained along a line (Fig. 4b), we find that the spectra at some points match that of P3HT or of PCBM. At some other points, the spectra however did not yield any of the two components. The location of HOMO and LUMO levels along the line has been plotted in Fig. 4c. The levels corresponding to points between 5 and 15 appeared to match that of PCBM; the energies of point 24 onwards similarly meant that the STM tip during those measurements was on P3HT. There are however other regions such as 1-4 and more clearly in between 16-23; the energy levels at these regions did not match with either of two materials. The HOMO and LUMO levels in these regions however remained correlated, that is, both the levels shifted in unison and also in the same the direction. The results hence imply that the regions refer to interface between the two materials and the energy levels at these regions underwent a bending. That is, P3HT influenced energy-levels of PCBM in a domain and vice-versa.

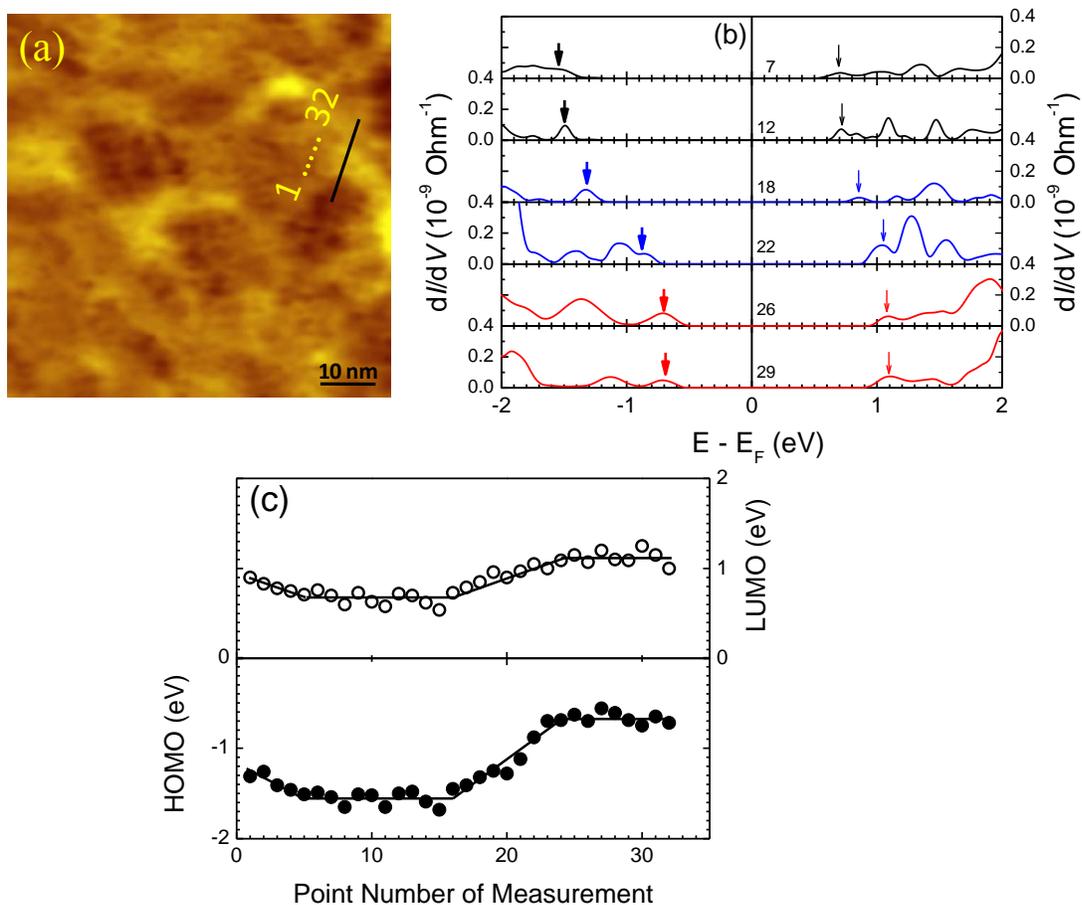

FIG. 4. (a) STM topography of a P3HT:PCBM BHJ. (b) d$I$/d$V$ spectra recorded at different points on the line drawn on the topography. The thick and thin-arrows indicate HOMO and LUMO levels, respectively. (c) HOMO and LUMO energies as derived from all the spectra versus point-number as marked on the topography; the lines are to guide the eye.

We then recorded d$I$/d$V$ images of the BHJ at different voltages. A bright color represented a higher DOS, whereas dimmed-sections symbolized for low DOS. Some images at representative voltages have been presented in Figs. 5a-f. To analyze the images, it is imperative to do so in conjunction with the energy level diagram of the BHJ (Fig. 3). In the negative substrate-voltage that probes HOMO of the materials, if the bias was above the HOMO level of P3HT, none of the materials should be seen in d$I$/d$V$ images, since little DOS would be available to be imaged. As (magnitude of) voltages becomes close to the HOMO of P3HT, domains of the donor materials become visible (Fig. 5a). When magnitude of the bias is increased further, the domains become brighter implying a higher DOS; the domains appeared to have grown with edge having a different color than that inside the domain implying different intensity of DOS at the edge (Fig. 5b). Such region in effect depicts the interfaces that have HOMO energies intermediate to that of the two materials (Fig. 2c) or bending in energy levels as shown in Fig. 4c. At still higher negative bias (below HOMO of PCBM) both the materials have appeared in the d$I$/d$V$ image (Fig. 5c). In the positive substrate-voltages, which probe the LUMO levels, there can be three regions as well: (i) close to the LUMO level of PCBM, (ii) in-between the two LUMOs, and (ii) above the LUMO of P3HT. The images accordingly bring out (i) domains of PCBM, (ii) domains of PCBM along with the interface region, and (iii) domains of both the materials, as presented in Figs. 5d-f, respectively.

If we look at the images closely and compare, they appeared complimentary to each other. That is, brighter domains in one image became dimmer in the other, since the same region was probed at all the voltages. For example, in Fig. 5(b), two domains have been marked by circles. The larger circle highlights a dimmer domain that has a low DOS and thus represents PCBM section, since HOMO energies are probed at a negative substrate-voltage. The brighter section similarly must hence symbolize P3HT regions that have HOMO energies lesser in magnitude than the PCBM's. Such a bright domain representing P3HT has been marked by a smaller circle. If these domains are searched in Fig. 5(e), at which PCBM regions should appear brighter since LUMO levels are probed here, we could clearly locate those P3HT and PCBM regions. They however have a contrasting brightness; the PCBM section (larger circle) appeared bright, whereas the P3HT section (smaller circle) turned dim. Shape and size of the domains remained identical. That is, the bias-dependent d*I*/d*V* images can clearly bring out nano-domains in a BHJ, if the materials have different energy levels. The bias-dependent d*I*/d*V* images can moreover bring out interface regions between the two materials as being elaborated in the following discussion.

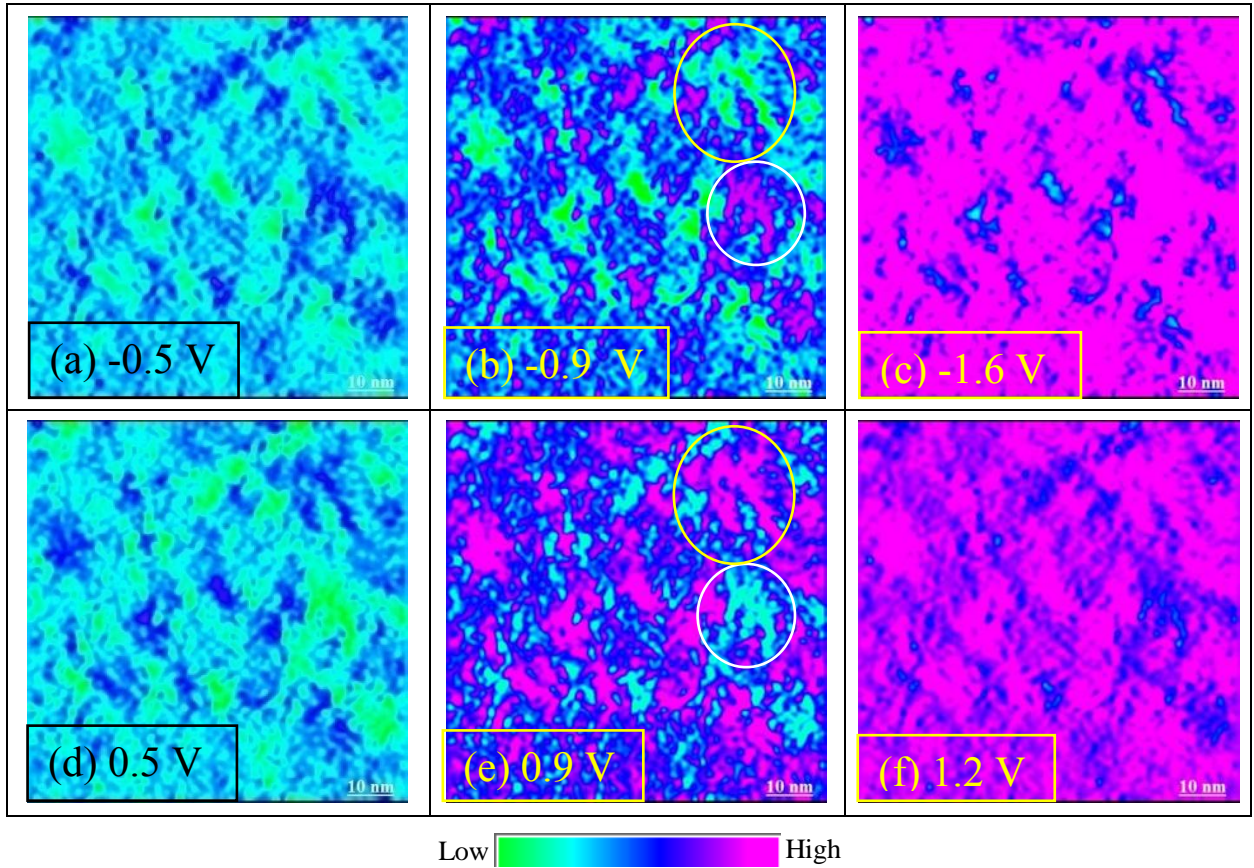

FIG. 5. d*I*/d*V* images of a P3HT:PCBM BHJ ultrathin-film recorded at different biases. At a negative substrate voltage, DOS of HOMO levels are probed; at positive voltage the DOS of LUMO levels are similarly probed. In (b) and (e), a large and a small circle represent a PCBM and a P3HT region, respectively.

We have thoroughly studied the images in sequence for better viewing of bias dependent d*I*/d*V* images. Since we recorded such images at many voltages in both bias-directions, we have clubbed the images to run as a function of voltage. The two "videos" for each bias-directions, as uploaded in the Supplemental Material [20], clearly allowed us to view the domains, followed by growth of the domains extending to the interface, and finally both the materials in thin-film BHJs. When the domains were growing, the color at the interface has been unalike of the two materials' domains, since the interface region possessed different energies than that of the components. The record of bias-dependent d$I$/d$V$ images through STS is in fact a unique method in viewing nano-domains of a BHJ and also the interface regions between two materials, *energetically*.

In conclusion, we have recorded d$I$/d$V$ images of the well-studied P3HT:PCBM BHJ at different voltages. To the best of our knowledge, this is the debut report on bias-dependent d$I$/d$V$ imaging of a BHJ that can be applied to any heterojunctions. Images of the BHJ elegantly allowed us to view the domains of P3HT and PCBM. From d$I$/d$V$ spectra of the BHJ, distribution in energy levels of the components brought out energy-level-bending at the interface. The voltage-dependent d$I$/d$V$ images inferred that the interface between the materials possessed energies different from the components. The images in addition allowed us to view the interface region with bending of HOMO (and LUMO) levels of the components towards each other. From the DOS spectra, we could moreover form energy diagram of the heterojunction as seen by charge carriers in BHJ films.

JC Bose National Fellowship (SB/S2/JCB-001/2016) and CSIR Junior Research Fellowship Number 09/080(1042)/2017-EMR-I (Roll No. 523509) are acknowledged.